\documentclass[aps,preprint,showpacs]{revtex4}
\usepackage{amsmath,latexsym,epsfig,times}

\begin{document}

\title{
Quantum discord for general X and CS  states:
A piecewise-analytic-numerical formula 
}

\author{
M.~A.~Yurischev
}
\email{yur@itp.ac.ru} 
\address{ 
Institute of Problems of Chemical Physics of the Russian Academy of Sciences,
142432 Chernogolovka, Moscow Region, Russia
}



\begin{abstract}
Quantum discord is a function of density-matrix elements (and through them,
e.~g., of temperature, applied fields, time, and so forth).
The domain of such a function in the case of two-qubit system with X or
centrosymmetric (CS) density matrix can consist at most of three subdomains:
two ones, where the quantum discord is expressed in closed analytical
forms ($Q_0$ and $Q_{\pi/2}$), and an intermediate subdomain in which for determining
the quantum discord $Q_\theta$ it is required to solve numerically a one-dimensional
minimization problem to find the optimal measurement angle $\theta\in(0,\pi/2)$.
Exact equations for determining the boundaries between these subdomains are obtained
and solved for a number of models.
The $Q_\theta$ subdomains are discovered in the anisotropic spin dimers in external field.
On the other hand, coinciding boundaries and therefore sudden transitions between optimal
measurement angles $\theta=\pi/2$ and $\theta=0$ are observed in dynamics
of spin currying particles in closed nanopore and also in phase flip channels.
In latter cases the solutions are entirely analytical.
\end{abstract}

\medskip 

\maketitle 

\section{Introduction}
\label{sec:Intro}
Quantum correlation theory is one of the most rapid developed direction
in modern physics.
This theory is important for the new technologies related to an utilization
of quantum information processing.

Earlier, only the quantum entanglement was attributed to quantum correlations
\cite{AFOV08,HHHH09}.
The quantum entanglement exists in non-separable states of a system. 
However in the last years one came to a conclusion that the quantum
correlations can be present in mixed {\em separable\/} states, i.~e.,
when the quantum entanglement is absent.
As a measure of total purely quantum correlations one takes now the
quantum discord \cite{CMS11,MBCPV12,AFY14}.
In the ground of discord conception lies the idea of measurements
performed on the system or its parts and extracting with their help a
maximum amount of information.

Due to necessary to solve the optimization problem, the evaluation of
quantum correlations, especially discord, is extremely hard.
If for the two-qubit systems the quantum entanglement has been obtained for
the arbitrary density matrices \cite{W98}, the analytical formulas for the
quantum discord were proposed only for X or CS states
\cite{Luo08,ARA10,FWBAC10,LWF11,DWZ11,Y13}.
In an X matrix, nonzero entries may belong only to the main diagonal and anti-diagonal.
The $n\times n$ CS matrix is defined by the relations for its elements:
$a_{i,j}=a_{n-i+1,n-j+1}$.
Any CS matrix of fourth order takes the X form under the double Hadamard transformation
$H\otimes H$ ($H$ is the ordinary Hadamard matrix) \cite{Y13}.
This transformation belongs to the class of local unitary transformations and therefore
the quantum correlations do not change their values.

However, it was latter found \cite{LMXW11,CZYYO11,H13} that the formulas for
X states \cite{ARA10,FWBAC10,LWF11} are not always correct.
The reason is that the authors \cite{ARA10,FWBAC10,LWF11} believed (and this
was their error) that the optimal measurements are achieved only in the limiting
points, i.~e., at the angles $\theta=0$ or $\pi/2$ (see below).
But on the explicit  examples \cite{LMXW11,CZYYO11,H13} of X density matrices
it was proved that that the optimal measurements can take places at the
intermediate angles in the interval $(0,\pi/2)$.
Unfortunately, these examples with the density matrices do not clarify
the physical situation.

In the present paper we use the language of Hamiltonians.
We show that the domain of intermediate optimal angles can arise in the vicinity
of transition from the domain with optimal measurement angle $\theta=\pi/2$
to the domain with optimal angle $\theta=0$ (or vice versa).
We obtain the equations for the boundaries between these domains and investigate
their solutions for some models.

In the following sections we establish an isomorphism between the X density
matrices and Gibbs ones with XYZ Hamiltonians, prove the existence of intermediate
domains (with the optimal anglers $\theta\not=0, \pi/2$), derive the equations
for boundaries between different domains, and, finally, apply the developed
approach to a gas of spin-carrying particles in closed nanopore and to a
phase flip channel.

\section{X density matrices and XYZ spin dimers}
\label{sec:X-XYZ}
In the most general form, the X density matrix of two-qubit system is
given as
\begin{equation}
   \label{eq:rho-Xc}
   \rho
	 =\left(
      \begin{array}{cccc}
      a&0&0&u_1+iu_2\\
      0&b&v_1+iv_2&0\\
      0&v_1-iv_2&c&0\\
      u_1-iu_2&0&0&d
      \end{array}
   \right), 
\end{equation}
where $a+b+c+d=1$.
This matrix contains seven real parameters which must satisfy the inequalities
\begin{equation}
   \label{eq:ineqs}
   a, b, c, d\ge0,\quad ad\ge u_1^2+u_2^2, \quad bc\ge v_1^2+v_2^2.
\end{equation}
Decomposition of the matrix (\ref{eq:rho-Xc}) on the Pauli matrices
$\sigma_\alpha$ ($\alpha=x,y,z$) leads to its Bloch form
\begin{eqnarray}
   \label{eq:rhoXc-Bloch}
   \rho&=&{1\over4}\{ 1
   - [1 - 2(a+b)]\sigma_z\otimes1
   - [1 - 2(a+c)]1\otimes\sigma_z
   + 2(u_1 + v_1)\sigma_x\otimes\sigma_x 
   \nonumber\\
   &+& 2(v_1 - u_1)\sigma_y\otimes\sigma_y 
   + [1 - 2(b+c)]\sigma_z\otimes\sigma_z 
   - 2(u_2 - v_2)\sigma_x\otimes\sigma_y 
   \nonumber\\
   &-& 2(u_2 + v_2)\sigma_y\otimes\sigma_x\}
\end{eqnarray}
The expansion coefficients are the unary and binary correlation functions
and therefore seven parameters of density matrix are expressed through the
seven different correlators.

To clarify the reasons for appearance of intermediate optimal measurements,
consider the spin-1/2 dimers in thermal equilibrium state.
In accord with Eq.~(\ref{eq:rhoXc-Bloch}) we take the Hamiltonian in the form
\begin{equation}
   \label{eq:Hxxz-xy}
   {\cal H} = -{1\over2}(J_x\sigma_1^x\sigma_2^x 
   + J_y\sigma_1^y\sigma_2^y 
   + J_z\sigma_1^z\sigma_2^z
	 + B_1\sigma_1^z + B_2\sigma_2^z)
   + J_{xy}\sigma_1^x\sigma_2^y + J_{yx}\sigma_1^y\sigma_2^x,
\end{equation}
where $\sigma_i^\alpha$ is the $\alpha$-th Pauli matrix in the
$i$-th site, and $J_x, J_y, J_z, B_1, B_2, J_{xy}$, and $J_{yx}$ are
seven arbitrary real parameters.
The corresponding Gibbs density matrix
\begin{equation}
   \label{eq:rho-G}
   \rho={1\over Z}e^{-{\cal H}/T}
\end{equation}
($T$ is the temperature in energy units and $Z$ is the partition function)
has the seven-parameter X form.
This circumstance allows to come from the formal density-matrix language to
the physically clear picture of interactions in the system. 

The quantum entanglement and quantum discord are invariant under the local
unitary transformations \cite{AFOV08,HHHH09,CMS11,MBCPV12,AFY14}.
Thanks to this property, one can with the help of transformation
\begin{equation}
   \label{eq:U}
   U=e^{-i\varphi_1\sigma_z/2}\otimes e^{-i\varphi_2\sigma_z/2}
\end{equation}
reduce the seven-parameters density matrix (\ref{eq:rho-Xc}) to the
{\em real} five-parameters X form \cite{H13,CRC10}.
This provides with the angles
\begin{equation}
   \label{eq:phi12}
   \varphi_{1,2}={1\over2}(\arctan{u_2\over u_1}\pm\arctan{v_2\over v_1}).
\end{equation}
After this, the density matrix (\ref{eq:rho-Xc}) takes the form
\begin{equation}
   \label{eq:rho-Xr}
   \rho
	 =\left(
      \begin{array}{cccc}
      a&0&0&u\\
      0&b&v&0\\
      0&v&c&0\\
      u&0&0&d
      \end{array}
   \right), 
\end{equation}
where 
\begin{equation}
   \label{eq:u}
   u=u_1\cos(\arctan{u_2\over u_1}) + u_2\sin(\arctan{u_2\over u_1}), 
\end{equation}
\begin{equation}
   \label{eq:v}
   v=v_1\cos(\arctan{v_2\over v_1}) + v_2\sin(\arctan{v_2\over v_1}). 
\end{equation}
Moreover, with the help of local rotations again around the $z$ axis, it is
not difficult to obtain also the non-negative off-diagonal elements of the
X matrix (\ref{eq:rho-Xr}).
Indeed, the local unitary transformation
\begin{equation}
   \label{eq:U1}
   U_1=e^{-i{\pi\over4}\sigma_z}\otimes e^{i{\pi\over4}\sigma_z}
   =\left(
      \begin{array}{cccc}
      i&&&\\
      &1&&\\
      &&1&\\
      &&&-i
      \end{array}
   \right), 
\end{equation}
leads to a change of sign for $u$:
\begin{equation}
   \label{eq:rho-U1}
   U_1\rho U_1^+
	 =\left(
      \begin{array}{rrrr}
      a&0&0&-u\\
      0&b&v&0\\
      0&v&c&0\\
      -u&0&0&d
      \end{array}
   \right). 
\end{equation}
Similarly, the local transformation
\begin{equation}
   \label{eq:U2}
   U_2=e^{i{\pi\over4}\sigma_z}\otimes e^{-i{\pi\over4}\sigma_z}
   =\left(
      \begin{array}{cccc}
      1&&&\\
      &i&&\\
      &&-i&\\
      &&&1
      \end{array}
   \right), 
\end{equation}
selectively acts on the sign of $v$:
\begin{equation}
   \label{eq:rho-U2}
   U_2\rho U_2^+
	 =\left(
      \begin{array}{rrrr}
      a&0&0&u\\
      0&b&-v&0\\
      0&-v&c&0\\
      u&0&0&d
      \end{array}
   \right). 
\end{equation}
Thus, after transformation of an X matrix to the real form we may simply enclose
the off-diagonal elements in the modul symbols:
\begin{equation}
   \label{eq:rho-Xm}
   \rho
	 =\left(
      \begin{array}{cccc}
      a&0&0&|u|\\
      0&b&|v|&0\\
      0&|v|&c&0\\
      |u|&0&0&d
      \end{array}
   \right). 
\end{equation}
This operation does not influence on the value of quantum correlations
in a system.

It is not difficult to understand that all peculiarities of discord behavior
in the states with seven-parameters density matrix (\ref{eq:rho-Xc}) can be
described by the thermal density matrix of XYZ dimer in inhomogeneous fields
$B_1$ and $B_2$ without crossing terms $J_{xy}$ and $J_{yx}$,
\begin{equation}
   \label{eq:Hxyz}
   {\cal H} = -{1\over2}(J_x\sigma_1^x\sigma_2^x 
   + J_y\sigma_1^y\sigma_2^y 
   + J_z\sigma_1^z\sigma_2^z
	 + B_1\sigma_1^z + B_2\sigma_2^z).
\end{equation}
This model contains five independent parameters
$J_x, J_y, J_z, B_1$, and $B_2$.

\section{Three alternatives for the quantum discord}
\label{sec:3Q}
As mentioned above, the measurement operations lie in the ground of discord
notion.
Following the founders of discord conception \cite{OZ01,Z03} and their adherents
\cite{Luo08,ARA10,FWBAC10,LWF11,DWZ11} proposed the formulas for calculation of
quantum discord in two-qubit systems, we will consider here only the
projective measurements \cite{rem1}.
For the X state (\ref{eq:rhoXc-Bloch}) and XYZ dimer (\ref{eq:Hxyz}),
$z$-direction is, obviously, peculiar.
Therefore, the measurements can be reduced to projections which are characterized
by the polar ($\theta$) and azimuthal ($\phi$) angles relative to the $z$-axis.
It is important that in the case of real X density matrix with an additional
condition $uv\ge0$ the optimal measurements are achieved by $\cos2\phi=1$
\cite{H13,CRC10}.

Let the spins 1 and 2 of a dimer be the subsystems $A$ and $B$, respectively.
Denote the density matrix of total system as $\rho_{AB}\ (=\rho)$, and for
the reduced density matrices we will use the notations $\rho_A$ and $\rho_B$:
\begin{equation}
   \label{eq:rhoA}
   \rho_A
   ={\rm Tr}_B\rho_{AB}
	 =\left(
      \begin{array}{cc}
      a+b&0\\
      0&c+d
      \end{array}
   \right), 
\end{equation}
\begin{equation}
   \label{eq:rhoB}
   \rho_B
   ={\rm Tr}_A\rho_{AB}
	 =\left(
      \begin{array}{cc}
      a+c&0\\
      0&b+d
      \end{array}
   \right). 
\end{equation}
Quantum discord in general depends on which subsystem ($A$ or $B$) the measurements
are performed.
Let, for definiteness, the measured subsystem be $B$.
Then the quantum discord is given as
\begin{equation}
   \label{eq:Q}
   Q=S(\rho_B)-S(\rho_{AB})+\min_\theta S_{cond}(\theta).
\end{equation}
Here $S(\rho)=-{\rm Tr}\rho\ln\rho$ is the von Neumann entropy (in nats)
for corresponding state $\rho$:
\begin{equation}
   \label{eq:SB}
   S(\rho_B)=-(a+c)\ln(a+c)-(b+d)\ln(b+d),
\end{equation}
\begin{eqnarray}
   \label{eq:S}
   &&S(\rho_{AB})\equiv S
   \nonumber\\
	 &&=-{a+d+\sqrt{(a-d)^2+4u^2}\over2}\ln{a+d+\sqrt{(a-d)^2+4u^2}\over2}
   \nonumber\\
	 &&-{a+d-\sqrt{(a-d)^2+4u^2}\over2}\ln{a+d-\sqrt{(a-d)^2+4u^2}\over2}
   \nonumber\\
	 &&-{b+c+\sqrt{(b-c)^2+4v^2}\over2}\ln{b+c+\sqrt{(b-c)^2+4v^2}\over2}
   \nonumber\\
	 &&-{b+c-\sqrt{(b-c)^2+4v^2}\over2}\ln{b+c-\sqrt{(b-c)^2+4v^2}\over2}.
\end{eqnarray}
Taking into account that $\cos2\phi=1$, the quantum conditional entropy
of subsystem $A$ is given as \cite{H13} 
\begin{equation}
   \label{eq:Scond}
   S_{cond}(\theta)=\Lambda_1\ln\Lambda_1+\Lambda_2\ln\Lambda_2
	 -\sum_{i=1}^4\lambda_i\ln\lambda_i, 
\end{equation}
where
\begin{equation}
   \label{eq:Lam12}
   \Lambda_{1,2}={1\over2}[1\pm (a-b+c-d)\cos\theta],
\end{equation}
\begin{eqnarray}
   \label{eq:lam12}
   \lambda_{1,2}&=&{1\over4}\lbrack\!\lbrack1+(a-b+c-d)\cos\theta
   \nonumber\\
	 &\pm&\{[a+b-c-d+(a-b-c+d)\cos\theta]^2+4(|u|+|v|)^2\sin^2\theta\}^{1/2}\rbrack\!\rbrack,
\end{eqnarray}
\begin{eqnarray}
   \label{eq:lam34}
   \lambda_{3,4}&=&{1\over4}\lbrack\!\lbrack1-(a-b+c-d)\cos\theta
   \nonumber\\
	 &\pm&\{[a+b-c-d-(a-b-c+d)\cos\theta]^2+4(|u|+|v|)^2\sin^2\theta\}^{1/2}\rbrack\!\rbrack.
\end{eqnarray}
The conditional entropy $S_{cond}(\theta)$ is a continuous and differentiable function
of its argument $\theta$. 

\begin{figure}[t]
\begin{center}
\epsfig{file=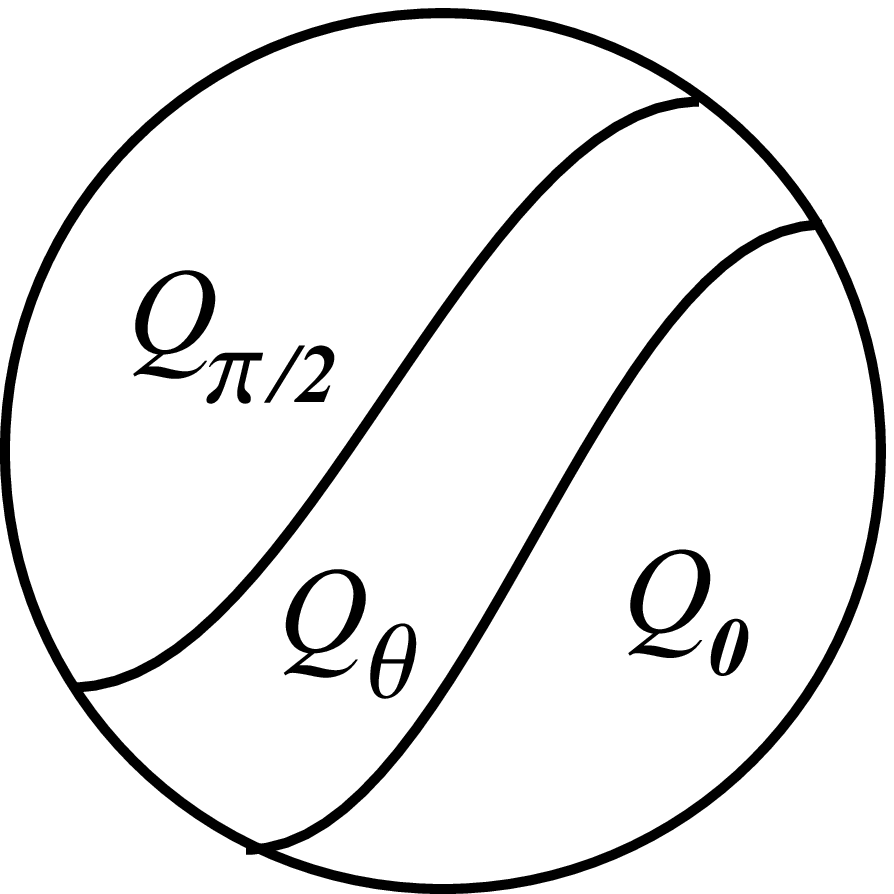,width=6.8cm}
\caption{A fragment of phase diagram with three possible domains for the 
quantum discord.}
\label{fig:ph-d}
\end{center}
\end{figure}

Expressions (\ref{eq:SB})-(\ref{eq:lam34}) allows to define the measurement-dependent
discord as
\begin{equation}
   \label{eq:Q-theta}
   Q(\theta)=S(\rho_B)-S(\rho_{AB})+S_{cond}(\theta),
\end{equation}
where $\theta\in[0,\pi/2]$.
The absolute minimum of this discord can be, obviously, either on the bounds
($\theta=0,\pi/2$) or at the intermediate point $\theta\in(0,\pi/2)$.
As a result, there is a choice from three corresponding possibilities
for the quantum discord
\begin{equation}
   \label{eq:Q3}
   Q=\min\{Q_0, Q_\theta, Q_{\pi\over2}\}.
\end{equation}
This equation generalizes the earlier proposed one for the quantum discord
\cite{Luo08,ARA10,FWBAC10,LWF11,DWZ11}
\begin{equation}
   \label{eq:Q2}
   \tilde Q=\min\{Q_0, Q_{\pi\over2}\},
\end{equation}
i.~e., the optimal observable can be either $\sigma_x$ or $\sigma_z$.
In Fig.~\ref{fig:ph-d} we schematically represent the parameter domain of a system
with three possible subdomains for the discord.

From the expressions~(\ref{eq:SB})-(\ref{eq:Q-theta}), we have for the
discord branch $Q_0\equiv Q(0)$:
\begin{equation}
   \label{eq:Q0}
   Q_0=-S-a\ln a-b\ln b-c\ln c -d\ln d.
\end{equation}
By $\theta=\pi/2$ we obtain
\begin{eqnarray}
   \label{eq:Q1}
   &&Q_{\pi\over2}=-S-\ln2-(a+c)\ln(a+c)-(b+d)\ln(b+d)
   \nonumber\\
	 &&-{1+\sqrt{(a+b-c-d)^2+4(|u|+|v|)^2}\over2}\ln{1+\sqrt{(a+b-c-d)^2+4(|u|+|v|)^2}\over4}
   \nonumber\\
	 &&-{1-\sqrt{(a+b-c-d)^2+4(|u|+|v|)^2}\over2}\ln{1-\sqrt{(a+b-c-d)^2+4(|u|+|v|)^2}\over4}.
\end{eqnarray}
Thus, the branches $Q_0$ and $Q_{\pi/2}$ are expressed analytically,
and the branch $Q_\theta=\min\nolimits_{\theta\in(0,\pi/2)} Q(\theta)$,
if the intermediate minimum exists, should be found
from the numerical solution of one-dimensional minimization problem or from
the transcendental equation
\begin{equation}
   \label{eq:theta}
   S^\prime_{cond}(\theta)=0,
\end{equation}
where the derivative of conditional entropy with respect to $\theta$ is equal to
\begin{equation}
   \label{eq:ScondI}
   S^\prime_{cond}(\theta)=\Lambda^\prime_1(1+\ln\Lambda_1)
	 +\Lambda^\prime_2(1+\ln\Lambda_2)
	 -\sum_{i=1}^4\lambda^\prime_i(1+\ln\lambda_i)
\end{equation}
with
\begin{equation}
   \label{eq:Lam12I}
   \Lambda^\prime_{1,2}=\mp{1\over2}(a-b+c-d)\sin\theta,
\end{equation}
\begin{eqnarray}
   \label{eq:lam12I}
   &&\lambda^\prime_{1,2}={1\over4}\bigg[-(a-b+c-d)\sin\theta
   \nonumber\\
	 &&\pm\frac{[a+b-c-d+(a-b-c+d)\cos\theta][-(a-b-c+d)\sin\theta]+2(|u|+|v|)^2\sin2\theta}
	 {\sqrt{[a+b-c-d+(a-b-c+d)\cos\theta]^2+4(|u|+|v|)^2\sin^2\theta}}\bigg],\quad
\end{eqnarray}
\begin{eqnarray}
   \label{eq:lam34I}
   &&\lambda^\prime_{3,4}={1\over4}\bigg[(a-b+c-d)\sin\theta
   \nonumber\\
	 &&\pm\frac{[a+b-c-d-(a-b-c+d)\cos\theta](a-b-c+d)\sin\theta+2(|u|+|v|)^2\sin2\theta}
	 {\sqrt{[a+b-c-d-(a-b-c+d)\cos\theta]^2+4(|u|+|v|)^2\sin^2\theta}}\bigg].
\end{eqnarray}
  
All three possible variants for the quantum discord ($Q_0$, $Q_{\pi/2}$, and $Q_\theta$)
can really exist in physical systems.
In the case when $a=b$ and $b=c$ (for example, the thermal discord for XYZ
dimers in absence of external field) the conditional entropy minimum
is achieved always at the bound points \cite{Luo08}.
The authors \cite{ARA10,FWBAC10,LWF11,DWZ11} have then extended the limiting possibilities
to the more general X states.
However, this is wrong; the minimum can take place at inner points.

\begin{figure}[t]
\begin{center}
\input{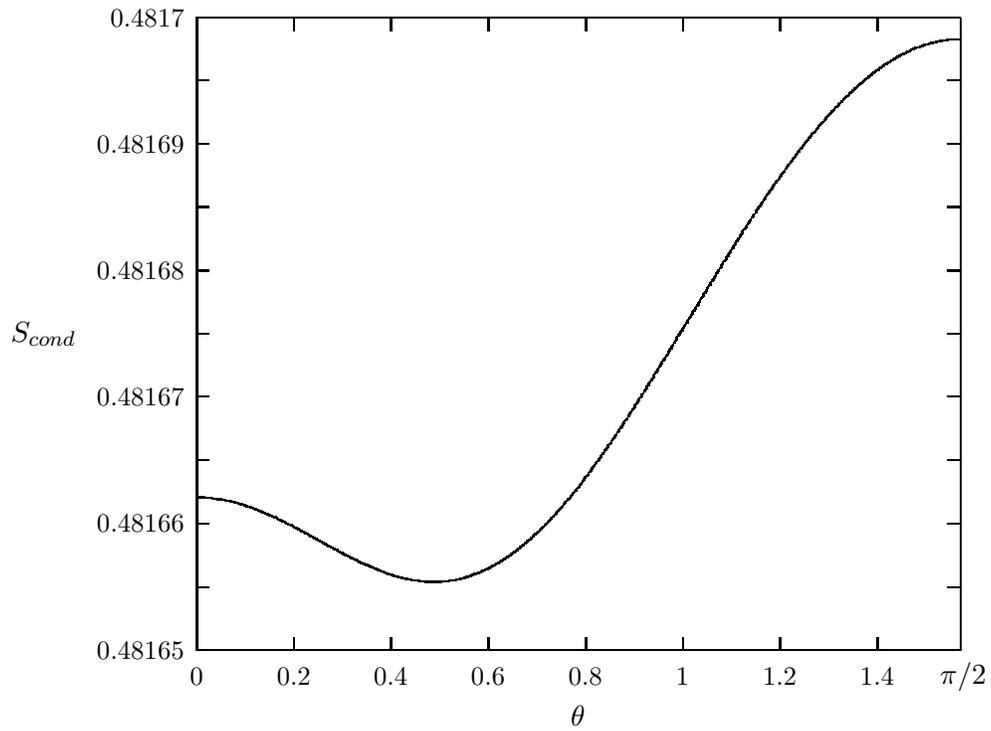}
\caption{Quantum conditional entropy $S_{cond}$ as a function of measured angle
$\theta$ for the state (\ref{eq:rho-LMXW}).}
\label{fig:LMXW}
\end{center}
\end{figure}

Indeed, following the authors \cite{LMXW11}, consider the state
\begin{equation}
   \label{eq:rho-LMXW}
   \rho
	 =\left(
      \begin{array}{cccc}
      0.0783&0&0&0\\
      0&0.125&0.100&0\\
      0&0.100&0.125&0\\
      0&0&0&0.6717
      \end{array}
   \right). 
\end{equation}
Using Eqs.~(\ref{eq:Scond})-(\ref{eq:lam34}) we computed the function
$S_{cond}(\theta)$ for this state.
Its behavior is shown in Fig.~\ref{fig:LMXW}.
It is clear seen that the conditional entropy minimum is situated in the
intermediate region, namely at the angle $\theta=0.4883\approx28^\circ$.
The quantum discord value is on $6.7\times10^{-6}$ less than the value
with $\theta=0$. 

\begin{figure}[t]
\begin{center}
\input{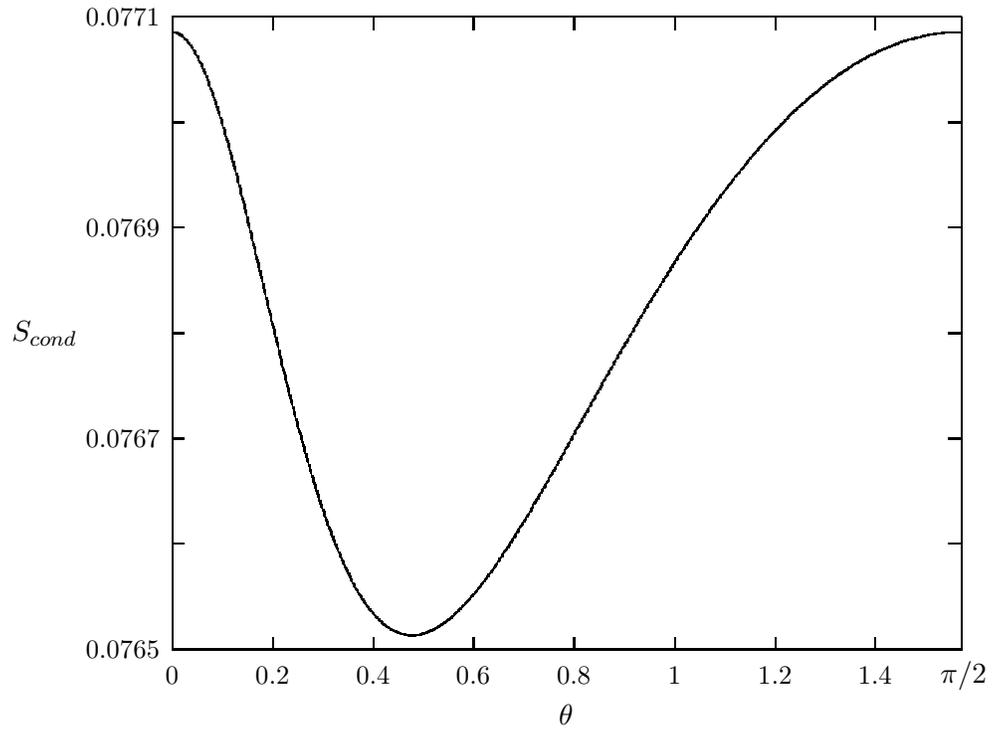}
\caption{Dependence $S_{cond}$ vs $\theta$ for the quantum state (\ref{eq:rho-H9}).}
\label{fig:H}
\end{center}
\end{figure}

In the paper \cite{H13}, other examples are given.
In particular, for the quantum state
\begin{equation}
   \label{eq:rho-H9}
   \rho
	 =\left(
      \begin{array}{cccc}
      0.021\,726&0&0&0.128\,057\\
      0&0.010\,288&0&0\\
      0&0&0.010\,288&0\\
      0.128\,057&0&0&0.957\,698
      \end{array}
   \right)
\end{equation}
the error for discord is here grater and equals $5.7\times10^{-4}$.
The behavior of $S_{cond}(\theta)$ for this state is depicted in Fig.~\ref{fig:H}.

These examples clearly show that the optimal measurement angles can really be
in the intermediate region $(0,\pi/2)$, i.~e., the optimal observables for quantum
discord can be not only the $\sigma_x$ or $\sigma_z$, but also their superposition.

For the real X state with constraint $|u+v|\ge|u-v|$,
the authors \cite{CZYYO11} have proven a theorem which guarantees that the optimal
observable is $\sigma_z$ if
\begin{equation}
   \label{eq:sz}
    (|u|+|v|)^2\le(a-b)(d-c)
\end{equation}
and $\sigma_x$ if
\begin{equation}
   \label{eq:sx}
    |\sqrt{ad}-\sqrt{bc}\le|u|+|v|.
\end{equation}
The theorem claims nothing for the region between these inequalities.
But in the case
\begin{equation}
   \label{eq:acbd}
    ac=bd
\end{equation}
the inequalities (\ref{eq:sz}) and (\ref{eq:sx}) lead to absence of any intermediate
region \cite{PKF13}.
In particular, this is valid for the Bell-diagonal states because for them $a=d$ and $b=c$.

\section{Equations for the exact boundaries}
\label{sec:bound}
Start with a heuristic example.
Consider the XXZ dimer in an uniform external field.
Its Hamiltonian is written as
\begin{equation}
   \label{eq:Hxxz}
   {\cal H} = -{1\over2}J(\sigma_1^x\sigma_2^x 
   + \sigma_1^y\sigma_2^y 
   + \Delta\sigma_1^z\sigma_2^z)
	 - {1\over2}B(\sigma_1^z + \sigma_2^z),
\end{equation}
where $\Delta=J_z/J$ is the coupling anisotropy.
The Gibbs density matrix is equal to
\begin{equation}
   \label{eq:rho-xxz}
   \rho
	 =\left(
      \begin{array}{cccc}
      a&&&\\
      &b&v&\\
      &v&b&\\
      &&&d
      \end{array}
   \right), 
\end{equation}
where
\begin{equation}
   \label{eq:abdv}
   a={1\over Z}e^{(J_z/2+B)/T},\quad
   b={1\over Z}e^{-J_z/2T}\cosh{J\over T},\quad
   d={1\over Z}e^{(J_z/2-B)/T},\quad
   v={1\over Z}e^{-J_z/2T}\sinh{J\over T},
\end{equation}
and the partition function
\begin{equation}
   \label{eq:Zxxz}
   Z=2(e^{J_z/2T}\cosh{B\over T}+ e^{-J_z/2T}\cosh{J\over T}).
\end{equation}
These expressions allow to determine the parameters of the Hamiltonian
(\ref{eq:Hxxz}):
\begin{equation}
   \label{eq:JJzB}
   J={T\over2}\ln\frac{1+v/b}{1-v/b},\quad
   J_z={T\over2}\ln\frac{ad}{b^2-v^2},\quad
   B={T\over2}\ln{a\over d}.
\end{equation}
Returning to the examples from the previous section we find that the
state (\ref{eq:rho-LMXW}) is realized for the dimer (\ref{eq:Hxxz}) with the
parameters $J=1$, $J_z=1.017\approx1.02$, and $B=-0.98$ at the temperature
$T=1.1$.
These values are quite reasonable.

Will now for each choice of interaction constants $J$ and $J_z$ find, on
the plane temperature-field, the lines which defined by the condition
\begin{equation}
   \label{eq:Q0Q1}
   Q_0(T,B)=Q_{\pi\over2}(T,B).
\end{equation}
After this we will study the changes of curves $S_{cond}(\theta)$ in the
neighborhood of those lines.

\begin{figure}[t]
\begin{center}
\input{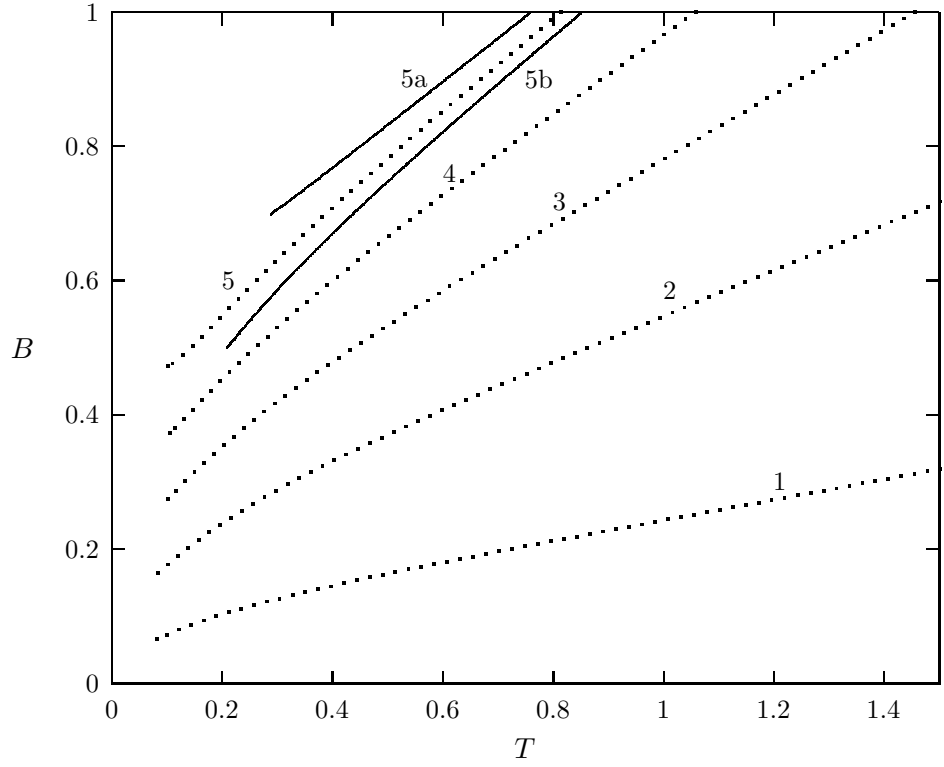}
\caption{Temperature-field diagram for the dimer (\ref{eq:Hxxz}).
The dashed lines are defined by Eq.~(\ref{eq:Q0Q1}) with $J=1$ and
$\Delta$: $1) 1.001$, 2)1.005, 3)1.01, 4)1.015, and 5)1.02.
Solid lines 5a and 5b correspond respectively to the $\pi/2$-
and 0-boundaries for the dimer with $\Delta=1.02$.}
\label{fig:TB}
\end{center}
\end{figure}

Using Eqs.~(\ref{eq:S}), (\ref{eq:Q0}), (\ref{eq:Q1}), (\ref{eq:abdv}),
and (\ref{eq:Zxxz}) we have numerically solved the transcendental
equation (\ref{eq:Q0Q1}) by $J=1$ and different values of $\Delta=J_z/J$.
The results are shown in Fig.~\ref{fig:TB} by dashed lines.

\begin{figure}[t]
\begin{center}
\input{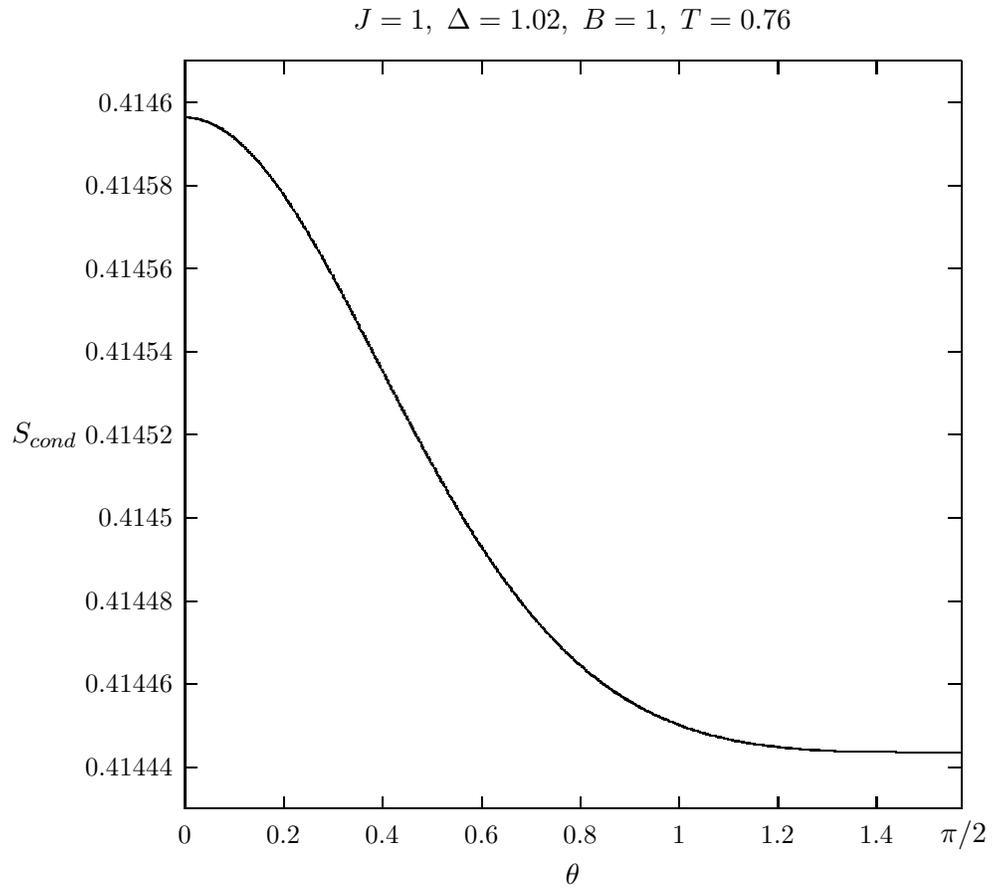}
\caption{
The form of $S_{cond}(\theta)$ for the XXZ dimer by
$J=1$, $\Delta=1.02$, and $B=1$.
The temperature $T=0.76$.}
\label{fig:Sa}
\end{center}
\end{figure}

\begin{figure}[t]
\begin{center}
\input{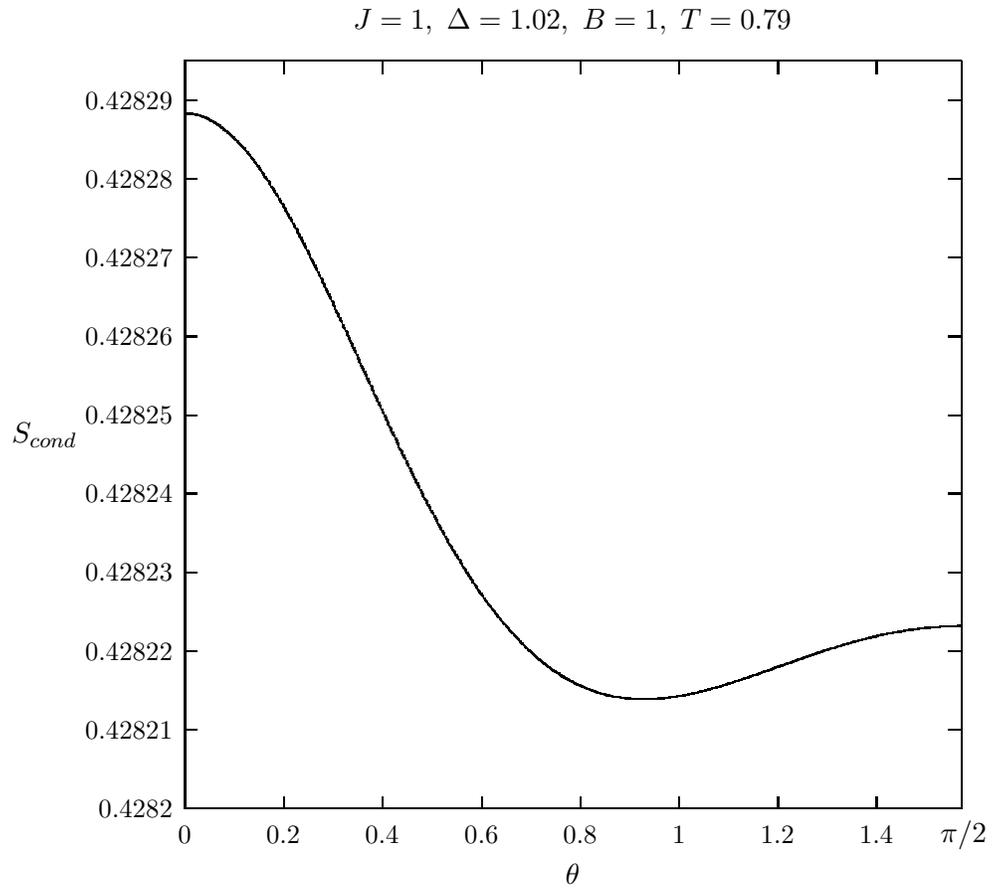}
\caption{The same as in Fig.~\ref{fig:Sa} but at $T=0.79$.}
\label{fig:Sb}
\end{center}
\end{figure}

\begin{figure}[t]
\begin{center}
\input{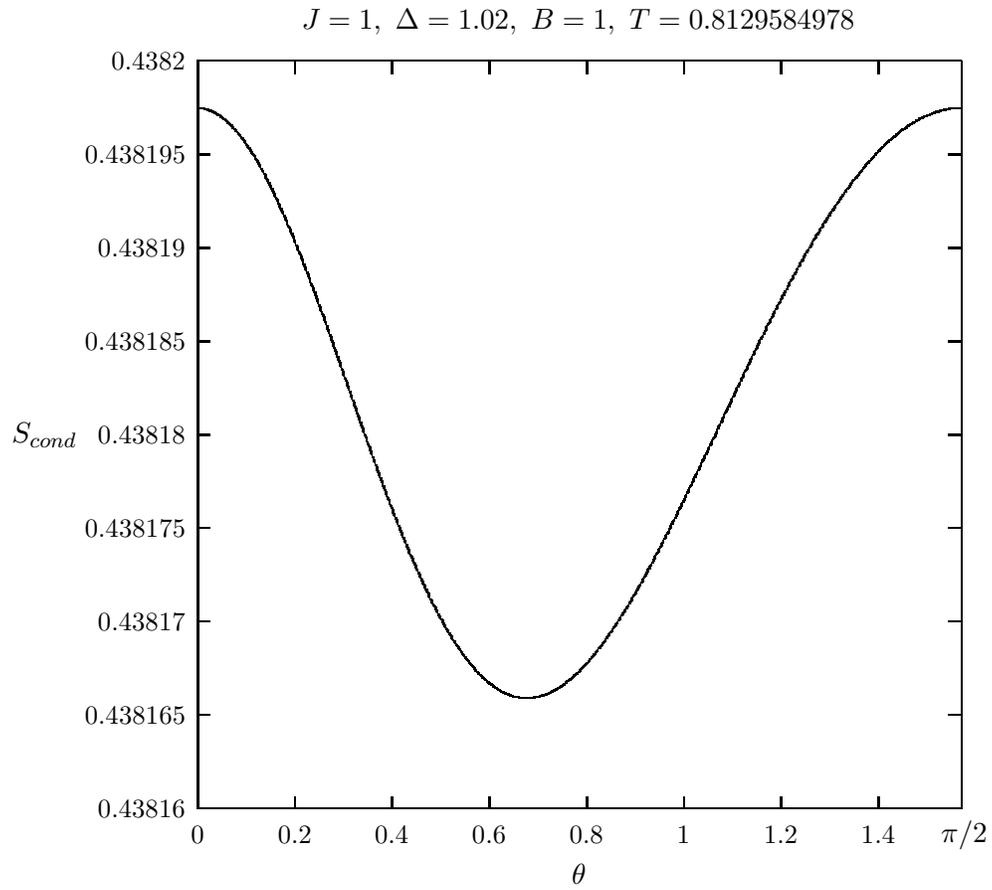}
\caption{The same as in Fig.~\ref{fig:Sa} but at the temperature $T_\times=0.81296$.}
\label{fig:Sc}
\end{center}
\end{figure}

Consider in detail, for example, the case $\Delta=1.02$ (the
5th line in Fig.~\ref{fig:TB}).
Let the external field is held fixed and equal to $B=1$.
Then the equality $Q_0=Q_{\pi/2}$ is satisfied at the temperature
$T_\times=0.81296$.
Study now the behavior of $S_{cond}(\theta)$ when the temperature varies.
If $T=0.76$, the minimum of $S_{cond}(\theta)$ is at $\theta=\pi/2$ (see
Fig.~\ref{fig:Sa}).
The angle $\theta=\pi/2$ is optimal for all lower temperatures.
When the temperature increases, it is appeared the minimum on the curve
$S_{cond}(\theta)$ inside the interval between 0 and $\pi/2$.
The minimum is clear seen by $T=0.79$ (Fig.~\ref{fig:Sb}).
Near the point $T_\times=0.81296$ the minimum achieves the maximal depth
(see Fig.~\ref{fig:Sc}).
With further increasing the temperature the minimum moves to the
bound $\theta=0$ (Fig.~\ref{fig:Sd}) and then it disappears at all
(Fig.~\ref{fig:Se}).
Optimal measurements undergo to the angle $\theta=0$.

Argue now that both lower and upper boundaries of the interval within which the
optimal angles lie between 0 and $\pi/2$ are exact,
i.~e., the intermediate minimum of $S_{cond}(\theta)$ suddenly appears and suddenly
disappears.
First of all, we note that the derivatives of function $S_{cond}(\theta)$
at $\theta=0$ and $\pi/2$ equal zero in general case:
$S^{\prime}_{cond}(0)=S^{\prime}_{cond}(\pi/2)=0$.
This is easy to check by direct calculations using
Eqs.~(\ref{eq:ScondI})-(\ref{eq:lam34I}).
Turn now again to the Figs.~\ref{fig:Sa}-\ref{fig:Se}.
By every fixed value of external field $B$ and for each value of temperature $T$
one can at any moment to say either exist the inside minimum or not.
For instance, when $T=0.76$ ($B=1$) the function $S_{cond}(\theta)$ is
concave at the point $\theta=\pi/2$ and therefore its second derivative
$S^{\prime\prime}_{cond}(\pi/2)<0$.
But when $T=0.79$ the conditional entropy has a local maximum at the same
bound point $\theta=\pi/2$ and therefore $S^{\prime\prime}_{cond}(\pi/2)>0$.
Hence, the bifurcation point (doubling the extremum) is determined by the
condition
\begin{equation}
   \label{eq:SII1}
   S^{\prime\prime}_{cond}(\pi/2)=0.
\end{equation}
Similarly for the other bound point $\theta=0$:
\begin{equation}
   \label{eq:SII0}
   S^{\prime\prime}_{cond}(0)=0.
\end{equation}
Using Eqs.~(\ref{eq:Scond})-(\ref{eq:lam34}) we obtain the second derivatives at
limiting points:
\begin{eqnarray}
   \label{eq:d2S0}
   S^{\prime\prime}_{cond}(0)&=&{1\over4}(a-b+c-d)\biggl(2\ln\frac{b+d}{a+c}+\ln{ac\over bd}\biggr)
   \nonumber\\
	 &&+{1\over4}(a-b-c+d)\ln{ad\over bc}-{1\over2}(|u|+|v|)^2\biggl({1\over a-c}\ln{a\over c}
	 +{1\over b-d}\ln{b\over d}\biggr)
\end{eqnarray}
and
\begin{eqnarray}
   \label{eq:d2S1}
   S^{\prime\prime}_{cond}(\pi/2)&=&(a-b+c-d)^2
   \nonumber\\
	 &&-{1\over2(1+r)}\bigl[a-b+c-d+{1\over r}(a+b-c-d)(a-b-c+d)\bigl]^2
   \nonumber\\
	 &&-{1\over2(1-r)}\bigl[a-b+c-d-{1\over r}(a+b-c-d)(a-b-c+d)\bigl]^2
   \nonumber\\
	 &&+{1\over2r}\{(a-b-c+d)^2[1-{1\over r^2}(a+b-c-d)^2]
	 -4(|u|+|v|)^2\}\ln\frac{1-r}{1+r},
\end{eqnarray}
where
\begin{equation}
   \label{eq:r}
   r=[(a+b-c-d)^2+4(|u|+|v|)^2]^{1/2}.
\end{equation}
The relations (\ref{eq:SII0})-(\ref{eq:r}) are the boundary equations for the
crossover subdomain $Q_\theta$.

If the solutions of Eqs.~(\ref{eq:SII1}) and (\ref{eq:SII0}) are the same then
the intermediate domain $Q_\theta$ is absent and the quantum discord is given by
analytical expressions.
On the other hand, instead roughly conditions (\ref{eq:sz}) and (\ref{eq:sx}),
the inequalities
  $S^{\prime\prime}_{cond}(0)\le0$
and
	$ S^{\prime\prime}_{cond}(\pi/2)\le0$
define now the whole subdomains $Q_0$ and $Q_{\pi/2}$ respectively.

Numerical solution of Eqs.~(\ref{eq:SII0})-(\ref{eq:r}) for the XXZ dimer
shows that the boundaries are the lines going approximately parallel to the
dashed lines (see the lines 5a and 5b in Fig.~\ref{fig:TB}).
As a result, it is arisen a domain within which the optimal angles should
be found numerically.
Out of this domain we have analytical expressions for the quantum discord.
To not clutter up the Fig.~\ref{fig:TB}, we show a part of intermediate domain
only for the case $\Delta=1.02$.
By $B=1$, the temperature of $\pi/2$-boundary equals $T_{\pi/2}=0.76106$
and for the 0-boundary $T_0=0.85361$.
The middle of this interval equals 0.80734 which is near the point
$T_\times=0.81296$.
The relative width of this interval is 11.5\%.

\begin{figure}[t]
\begin{center}
\input{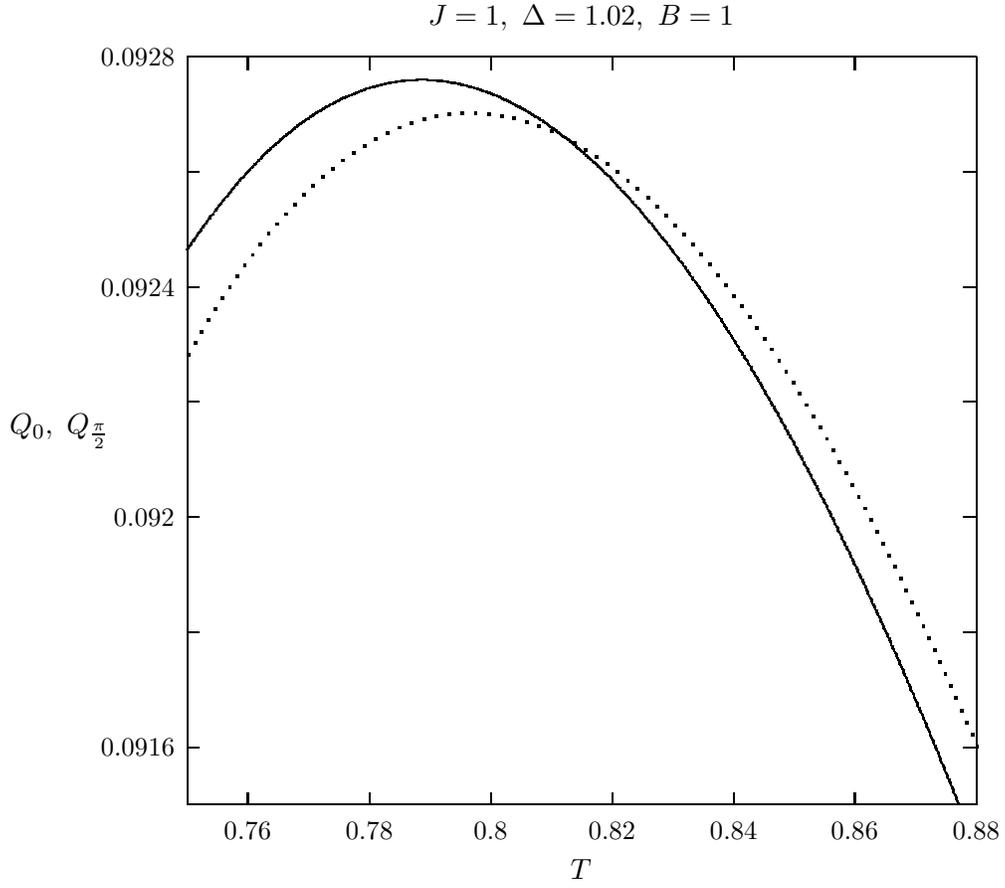}
\caption{Temperature behavior of $Q_0$ (solid line) and $Q_{\pi/2}$
(dashed line) for the XXZ dimer with parameters $J=1$, $J_z=1.02$, and $B=1$.}
\label{fig:Q0Q1}
\end{center}
\end{figure}

\begin{figure}[t]
\begin{center}
\input{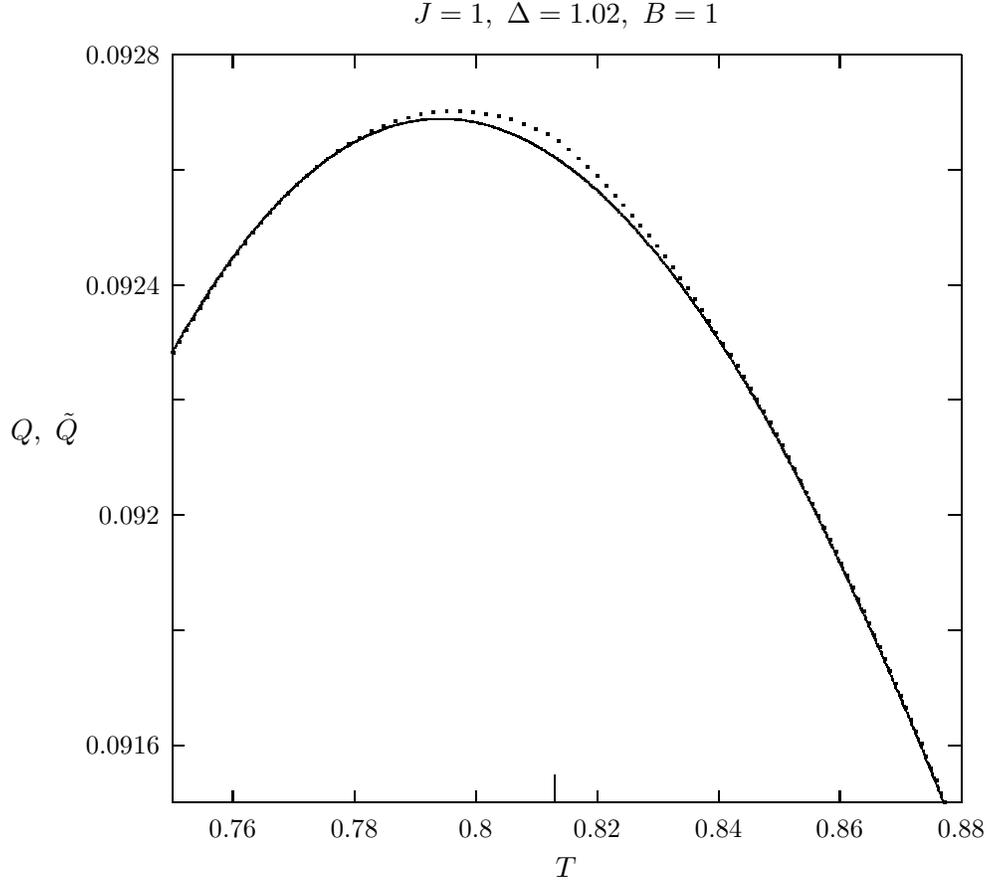}
\caption{Dependencies of the false discord $\tilde Q=\min\{Q_{\pi/2},Q_0\}$
(dashed line) and
the correct quantum discord $Q=\min\{Q_{\pi/2},Q_\theta,Q_0\}$ (solid line)
for the XXZ dimer with parameters $J=1$, $J_z=1.02$, and $B=1$.
The domain between the temperatures $T_{\pi/2}=0.76106$ and $T_0=0.85361$
corresponds to $Q_\theta$.
The longer bar marks the temperature $T_\times=0.81296$.}
\label{fig:QQ}
\end{center}
\end{figure}

\begin{figure}[t]
\begin{center}
\input{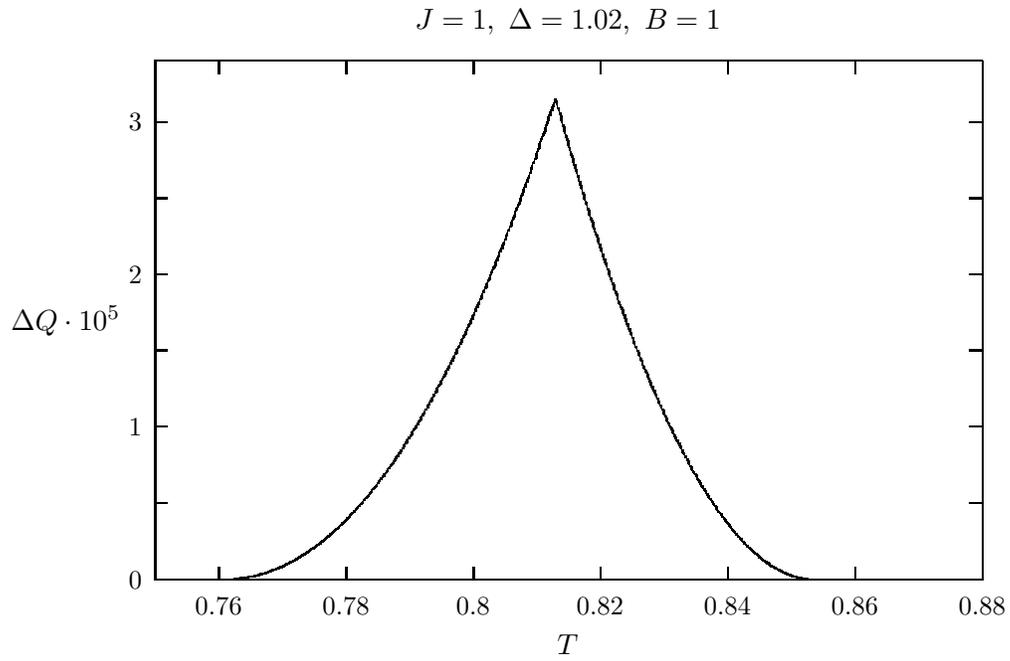}
\caption{Excess $\Delta Q={\tilde Q}-Q$ versus the temperature $T$
for the XXZ dimer with the parameters $J=1$, $J_z=1.02$, and $B=1$.
The excess is zero out the temperature interval $(0.76106,085361)$.}
\label{fig:DQ}
\end{center}
\end{figure}

Using the example of XXZ dimer with parameters $J=1$, $J_z=1.02$, and $B=1$
consider the thermal discord behavior by a transition from the domain
$Q_{\pi/2}$ into $Q_0$ one.
For this case, the functions $Q_0$ and $Q_{\pi/2}$ versus the temperature
are shown in Fig.~\ref{fig:Q0Q1}.
One can see that down to crossing point $T_\times=0.81296$ the
discord (${\tilde Q}$) as a
minimal value would be according Refs.~\cite{ARA10}-\cite{LWF11} equals
$Q_{\pi/2}$, and above the intersection point $T_\times$ equals $Q_0$.
If this would be valid, the discord ${\tilde Q}=\min\{Q_0,Q_{\pi/2}\}$
at the intersection point $T_\times$ would be not differentiable.
However, in reality the function is smooth.
This follows from the numerical solution of the task in the intermediate domain.
The results are shown in Fig.~\ref{fig:QQ}.
It is clearly seen that the smoothness occurs.
By this, the correction $\Delta Q={\tilde Q} -Q$ has a small value.
The behavior of $\Delta Q$ is shown in Fig.~\ref{fig:DQ}.
The curve has a cusp-like form.
The largest deviation $\Delta Q=3.1\times10^{-5}$ is about 0.03\%.

\begin{figure}[t]
\begin{center}
\input{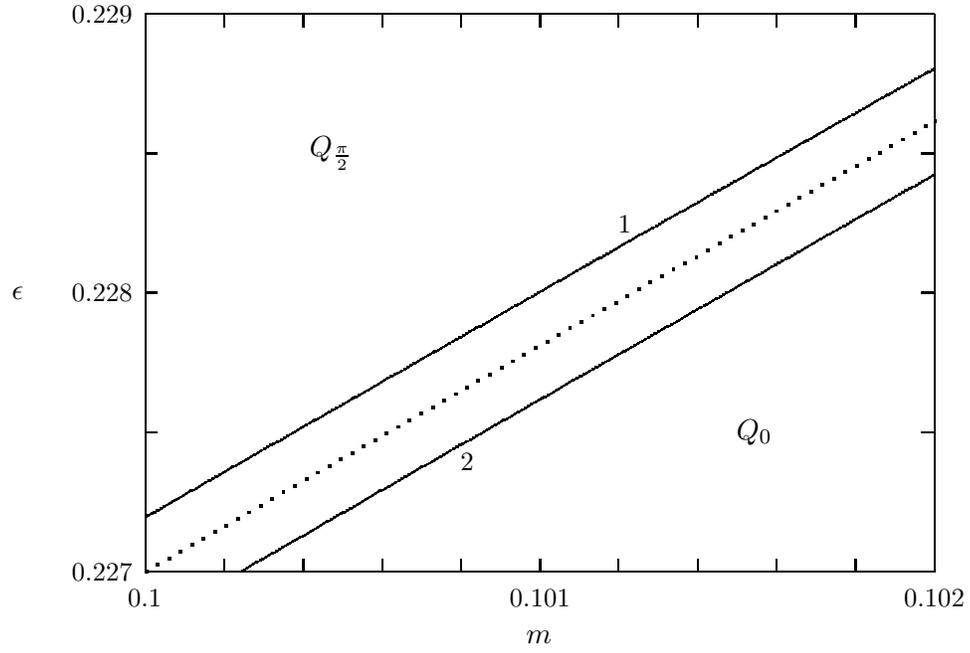}
\caption{Domains $Q_{\pi/2}$, $Q_0$, and (between them) $Q_\theta$ for the state
(\ref{eq:rho-me}).
Dotted line corresponds to the condition $Q_{\pi/2}=Q_0$.
Solid lines 1 and 2 are the $\pi/2$- and 0-boundaries respectively.}
\label{fig:zme}
\end{center}
\end{figure}

A two-parameter family of X states
\begin{equation}
   \label{eq:rho-me}
   \rho=\left(
      \begin{array}{cccc}
      \epsilon/2&0&0&\epsilon/2\\
      0&(1-\epsilon)m&0&0\\
      0&0&(1-\epsilon)(1-m)&0\\
      \epsilon/2&0&0&\epsilon/2
      \end{array}
   \right),
\end{equation}
was considered in Ref.~\cite{CZYYO11}.
Using Eqs.~(\ref{eq:SII1})-(\ref{eq:r}) we calculated the lower and upper boundaries
for the state (\ref{eq:rho-me}).
The results are presented in Fig.~\ref{fig:zme}.
Note that the sufficient conditions (\ref{eq:sz}) and (\ref{eq:sx}) for the $Q_0$
and $Q_{\pi/2}$ domains give \cite{CZYYO11}
\begin{equation}
   \label{eq:e0}
	 \epsilon\le\frac{2m(1-m)}{1+2m(1-m)}
\end{equation}
and
\begin{equation}
   \label{eq:e1}
	 \epsilon\ge\frac{\sqrt{m(1-m)}}{1+\sqrt{m(1-m)}}
\end{equation}
respectively.
Unfortunately, these boundaries rough too and lie far out the region of
Fig.~\ref{fig:zme}.

Notice, the boundaries may coincide between themselves.
Moreover, by coinciding boundaries, the cases can take places when there are no
smoothing for the quantum discord.
Such examples we discuss in the following sections.

\section{CS density matrices. Gas in nanopore}
\label{sec:pore}
As has been shown in Ref.~\cite{FKY12}, the reduced density matrix for any pair of
nuclear spins in a closed nanopore filled with a gas of spin-carrying
molecules has the CS form
\begin{equation}
   \label{eq:rho-pore}
   \rho=\left(
      \begin{array}{cccc}
      {1\over4}&{1\over2}p-iu&{1\over2}p-iu&q-r\\
      {1\over2}p+iu&{1\over4}&q+r&{1\over2}p+iu\\
      {1\over2}p+iu&q+r&{1\over4}&{1\over2}p+iu\\
      q-r&{1\over2}p-iu&{1\over2}p-iu&{1\over4}
      \end{array}
   \right),
\end{equation}
where 
\begin{eqnarray}
   \label{eq:pqru}
   &&p={1\over2}\tanh{\beta\over2}\cos^{N-1}(at),
   \nonumber\\
   &&q={1\over8}\tanh^2{\beta\over2}[1+\cos^{N-2}(2at)],
   \nonumber\\
   &&r={1\over8}\tanh^2{\beta\over2}[1-\cos^{N-2}(2at)],
   \\
   &&u={1\over4}\tanh{\beta\over2}\cos^{N-2}(at)\sin(at).
   \nonumber
\end{eqnarray}
(One should not confuse the defined here quantities $u$ and $r$ with the same denoted
quantities from the previous sections.)
In relations (\ref{eq:pqru}), $N$ is the number of particles confined in a nanopore,
$\beta$ is the inverse dimensionless temperature, and $\alpha t$ is the
dimensionless time.

After transformation to the real X form, the density matrix (\ref{eq:rho-pore})
takes the form
\begin{equation}
   \label{eq:rho11-pore}
   \rho=\left(
      \begin{array}{cccc}
      {1\over4}+p+q&0&0&2u\sin2\varphi-r\cos2\varphi\\
      0&{1\over4}-q&r&0\\
      0&r&{1\over4}-q&0\\
      2u\sin2\varphi-r\cos2\varphi&0&0&{1\over4}-p+q
      \end{array}
   \right),
\end{equation}
where
\begin{equation}
   \label{eq:phi}
   \varphi=-{1\over2}\arctan(2u/r).
\end{equation}
Using the analytical formulas \cite{FWBAC10} which are valid in a supposition
that the quantum conditional entropy has minimums only at the bound points,
the author \cite{Y13} found the quantum discord for particles in nanopore.
Here we prove that in the case of nanopore the intermediate optimal angle domain
does not exist and therefore the results \cite{Y13} are correct.

\begin{figure}[t]
\begin{center}
\input{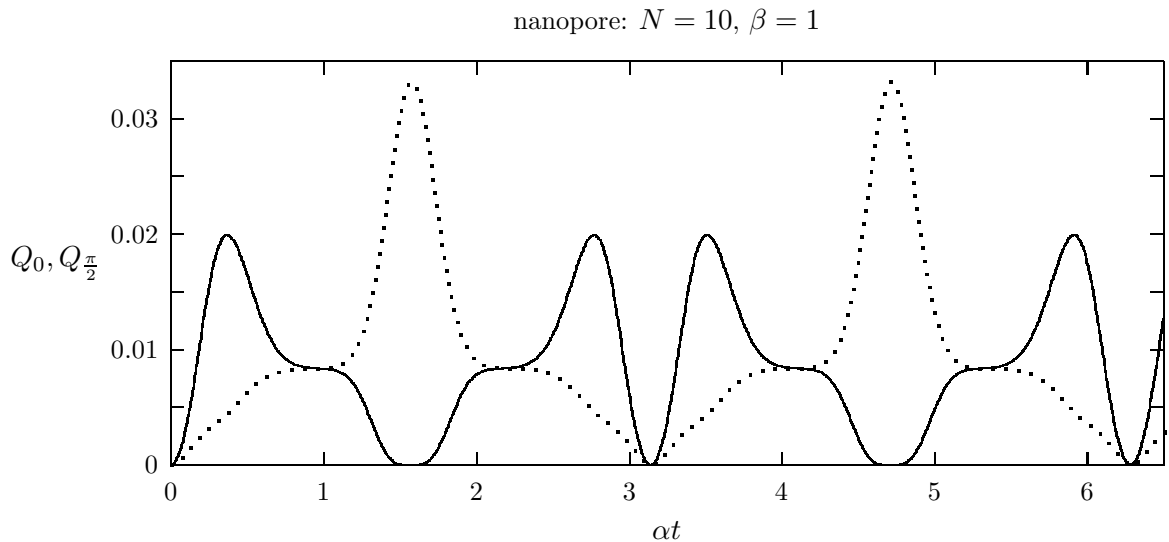}
\caption{Time dependencies of $Q_0$ (solid line) and $Q_{\pi/2}$ (dashed line)
in nanopore with $N=10$ and $\beta=1$.}
\label{fig:Qpore}
\end{center}
\end{figure}

\begin{figure}[t]
\begin{center}
\input{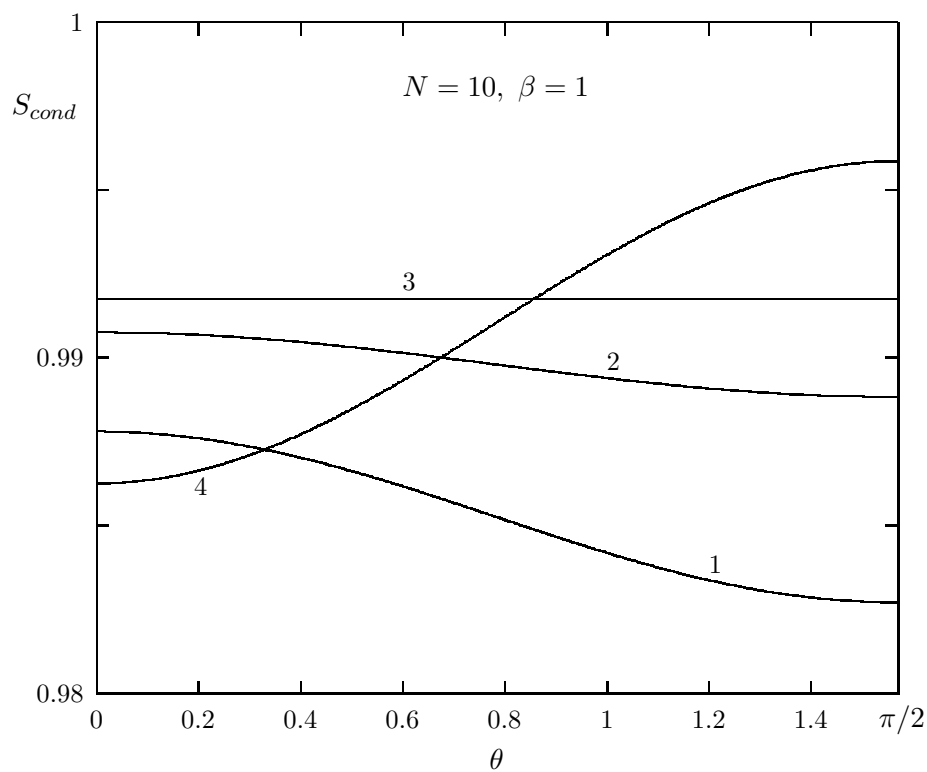}
\caption{Changes of $S_{cond}(\theta)$ form for a nanopore with $N=10$ and $\beta=1$.
The curves 1, 2, 3, and 4 correspond respectively to $\alpha t=0.6, 0.7, 0.98486$,
and 1.3.}
\label{fig:Spore}
\end{center}
\end{figure}

Applying the approach discussed above we study the behavior of functions
$Q_0(\alpha t)$ and $Q_{\pi/2}(\alpha t)$ by fixed values of $N$ and $\beta$.
The behavior of these quantities (in bits!) by $N=10$ and $\beta=1$ is shown in
Fig.~\ref{fig:Qpore}.
Both functions are periodic with the period equals $\pi$.
In the interval $0<\alpha t<\pi$, the curves are crossed at points
$\alpha t_1=0.98486$ and $\alpha t_2=2.15673$ (see Fig.~\ref{fig:Qpore}).
A solution of boundary equations (\ref{eq:SII0})-(\ref{eq:r}) shows that
in this case both boundaries coincide between themselves.
Thus, for the discussed system there are not domains where the conditional
entropy has the intermediate minimum.
Changes of the function $S_{cond}(\theta)$ in the vicinity of point $\alpha t_1$
is shown in Fig.~\ref{fig:Spore}.
When $t<t_1$, the conditional entropy has a minimum at $\theta=\pi/2$ (curves
1 and 2 in Fig.~\ref{fig:Spore}).
At $t=t_1$, the function $S_{cond}(\theta)$ becomes a straight line.
This is the point of indifference to a choice of measurement angle:
$\forall~\theta\in[0,\pi/2]$.
With further increasing $\alpha t$, the minimum of $S_{cond}(\theta)$ lies at
the second bound $\theta=0$.
At the another crossing point $\alpha t_2$ the quantum discord goes, vice
versa, from the branch $Q_0$ to $Q_{\pi/2}$, and so on.
Thus, the quantum discord $Q(\alpha t)=\min\{Q_0,Q_{\pi/2}\}$ is given here in
a closed analytical form.
Note that this function is smooth at the crossing points.

For odd $N$, the solution is another.
In this case there are no crossing points of curves $Q_0(\alpha t)$ and
$Q_{\pi/2}(\alpha t)$ and the quantum discord at any time is defined by
the branch $Q_{\pi/2}(\alpha t)$.

So, the quantum discord for nanopore is an analytical or piecewise analytical
function.

\section{Phase flip channel}
\label{sec:channel}
The authors \cite{LWF11} have considered the dynamics of quantum discord
under decoherence in a phase flip channel.
The problem is to calculate the quantum discord for the X matrix
\begin{eqnarray}
   \label{eq:eps}
   \varepsilon&=&{1\over4}[ 1
   + r\sigma_z\otimes1
   + s1\otimes\sigma_z
   + (1-p)^2c_1\sigma_x\otimes\sigma_x 
   \nonumber\\
   &+& (1-p)^2c_2\sigma_y\otimes\sigma_y 
   + c_3\sigma_z\otimes\sigma_z],
\end{eqnarray}
where $p=1-\exp(-\gamma t)$, $t$ is the time, and $\gamma$ is the phase damping
rate.
The authors \cite{LWF11} restricted themselves to the case where
\begin{equation}
   \label{eq:c2s}
   c_2=-c_3c_1,\qquad s=c_3r,\qquad -1\le c_3\le1,\qquad -1\le r\le1 .
\end{equation}
Expansion coefficients in Eq.~(\ref{eq:eps}) are related with the matrix elements as
\begin{eqnarray}
   \label{eq:a-v}
   &&a=(1+r+s+c_3)/4,
   \nonumber\\
   &&b=(1+r-s-c_3)/4,
   \nonumber\\
   &&c=(1-r+s-c_3)/4,
   \nonumber\\
   &&d=(1-r-s+c_3)/4,
   \\
   &&u=(1-p)^2(c_1-c_2)/4,
   \nonumber\\
   &&v=(1-p)^2(c_1+c_2)/4 .
   \nonumber
\end{eqnarray}
Notice that thanks to the relation $s=c_3r$ [see Eqs.~(\ref{eq:c2s})] the matrix
elements $a, b, c$, and $d$ satisfy the condition (\ref{eq:acbd}) and, as a result,
the $Q_\theta$ domain is absent here.

The authors \cite{LWF11} have established that the quantum discord in the model
under question does not change for a finite time interval.
This phenomenon they demonstrated for the channel with parameters $r=0.3$,
$s=0.15$, $c_1^2=4/5$, $c_2=-c_1/2$, and $c_3=1/2$.
It was found that in such a channel the sudden transition from the branch
$Q_{\pi/2}(p)$ to $Q_0(p)$ happens at $p_0=0.274$.
Moreover, at this point the quantum discord is continuous but its first derivative
is discontinuous (has a finite jump).
The authors \cite{LWF11} used the formulas for calculation of quantum discord without
intermediate measurement angles.

\begin{figure}[t]
\begin{center}
\input{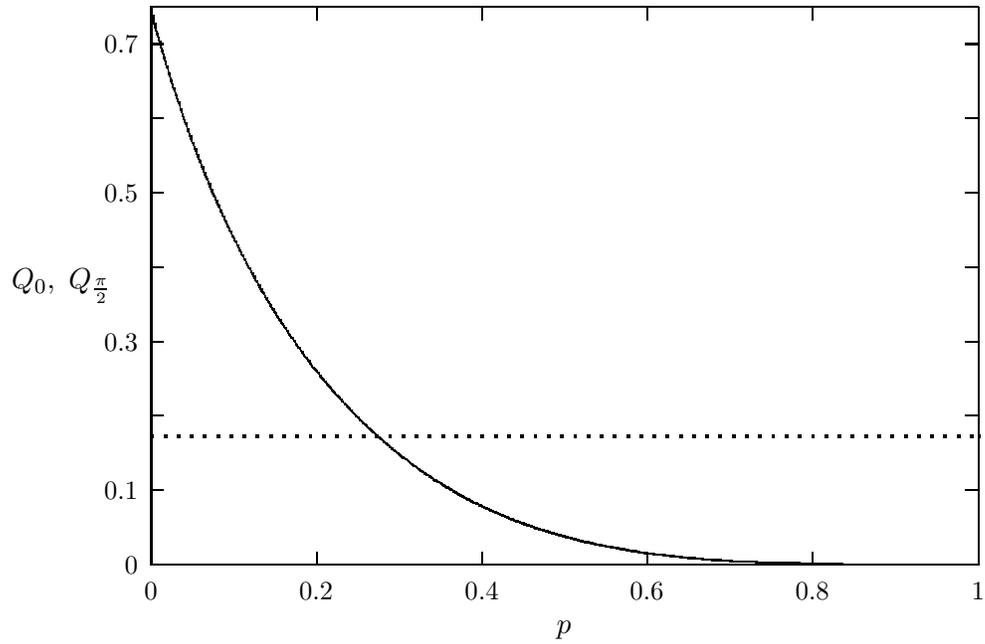}
\caption{$Q_0$ (solid line) and $Q_{\pi/2}$ (dashed line)
vs $p$ for a phase flip channel with parameters $r=0.3$, $s=0.15$, $c_1^2=4/5$,
$c_2=-c_1/2$, and $c_3=1/2$.
Crossing point of the lines is at $p_0=0.2743\ldots\ $.}
\label{fig:zq0q1c}
\end{center}
\end{figure}

\begin{figure}[t]
\begin{center}
\input{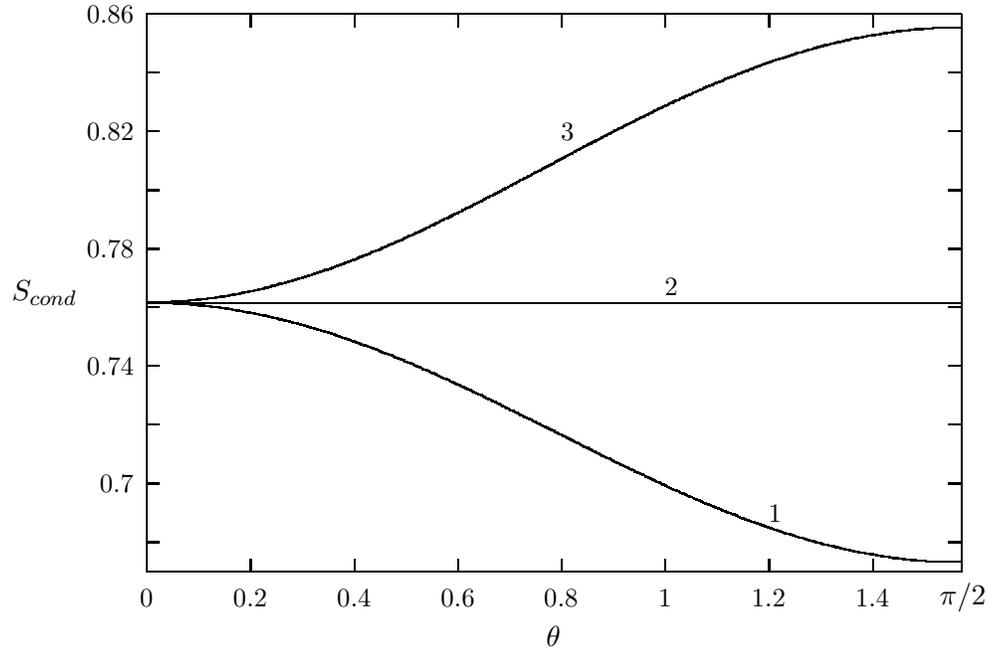}
\caption{$S_{cond}(\theta)$ for a phase flip channel
with parameters $r=0.3$, $s=0.15$, $c_1^2=4/5$, $c_2=-c_1/2$, and $c_3=1/2$.
The curves 1, 2, and 3 correspond respectively to $p=0.2$, 0.2743, and 0.4.}
\label{fig:zs2743}
\end{center}
\end{figure}

Using the approach developed, find the discord for the channel (\ref{eq:eps}) with
the above parameters.
First, we investigate the behavior of functions $Q_{\pi/2}(p)$ and $Q_0(p)$.
As seen in Fig.~\ref{fig:zq0q1c}, they (in bits!) have a crossing point at $p_0$.
Then, the solution of equations for the boundaries, Eqs.~(\ref{eq:SII1})-(\ref{eq:r}),
shows that the $\pi/2$- and 0-boundaries coincide between themselves, i.~e.,
the intermediate region is here reduced to a point.
The changes of $S_{cond}(\theta)$ form are presented in Fig.~\ref{fig:zs2743}.
It is seen that the transition $Q_{\pi/2}\rightarrow Q_0$ goes through the
straight line (no intermediate minimum occurs in the vicinity of point $p_0$).
So, in this example, the quantum discord $Q=\min\{Q_{\pi/2},Q_0\}$
is given in the closed analytical form.
It is a continuous but piecewise smooth function.

\section{Conclusions}
\label{sec:concl}
In the light of above, the calculation of quantum discord for general X
states can be reduced to the following steps.
At first one should transform the density matrix (\ref{eq:rho-Xc}) to
the real form, i.~e., calculate the quantities $u$ and $v$ using
Eqs.~(\ref{eq:u}) and (\ref{eq:v}).
Then one should solve equation $Q_0=Q_{\pi/2}$ and determine the
possible intersection points of branches $Q_0$ and $Q_{\pi/2}$.
After this one solves equations $S_{cond}^{\prime\prime}(0)=0$ and
$S_{cond}^{\prime\prime}(\pi/2)=0$ to find the boundaries
for the intermediate domain $Q_{\theta}$.
As a result the quantum discord is given as $Q=\min\{Q_0, Q_\theta, Q_{\pi/2}\}$.

The formula for calculation of quantum discord belongs to a
piecewise-defined type
\begin{equation}
   \label{eq:f}
   f(x)=
	 \begin{cases}
      F(x,a),\ x\in\Omega_a\\
	    F(a,b),\ x\in\Omega_b\\
	    \min_{\alpha\in(a,b)} F(x,\alpha),\ x\in\Omega_c
   \end{cases}.
\end{equation}
In other words, the domain of definition $\Omega$ of the function $f(x)$
consists of subdomains in which the function is given by closed
analytical expressions or it exits only in a numerical form.

In the case of CS density matrix $\rho_{CS}$ we, first of all, transform it
to the X form $\rho_X$ and then repeat the above steps for latter.
The transformation $R\rho_{CS}R=\rho_X$ is achieved with the help of
$R=H\otimes H$, where
\begin{equation}
   \label{eq:Had}
   H={1\over\sqrt{2}}\left(
      \begin{array}{rr}
      1&1\\
      1&-1
      \end{array}
   \right)
\end{equation}
is the Hadamard transform.

In this paper, we have shown also that the optimal intermediate
measurement angles $\theta\in(0,\pi/2)$ can occur in the transition
domain from the $Q_0$ to $Q_{\pi/2}$ or reversely.
The boundaries of this domain are exactly defined.
The corresponding equations for this boundaries have been found.
The boundaries may coincide between themselves and then the quantum
discord is evaluated analytically in the total domain of definition.

The domains $Q_\theta$ with the optimal intermediate angles $\theta\in(0,\pi/2)$
have been discovered in weakly-anisotropic spin dimers.
It has been shown also that the quantum discord for any pair of spin-carrying
particles confined in nanopore has everywhere the analytical representation.
Lastly, the quantum discord for a phase flip channel is continuous but has
the discontinuous first derivative at sudden transition point.

\section*{Acknowledgments}
\label{sec:akn}
The author thanks A.~I.~Zenchuk for valuable remarks.
The research was supported by the RFBR grants Nos.~13-03-12418 and 13-03-00017
and by the programs No.~8 of the Presidium of RAS and No.~14-042 of the Chemistry
and Material Science Department of RAS.


\newpage
\clearpage

\newpage
\clearpage

\newpage
\clearpage

\newpage
\clearpage

\newpage
\clearpage

\newpage
\clearpage

\newpage
\clearpage
\begin{figure}[t]
\begin{center}
\input{fig5d.tex}
\caption{The same as in Fig.~\ref{fig:Sa} but at $T=0.83$.}
\label{fig:Sd}
\end{center}
\end{figure}

\newpage
\clearpage
\begin{figure}[t]
\begin{center}
\input{fig5e.tex}
\caption{The same as in Fig.~\ref{fig:Sa} but at $T=0.85$.}
\label{fig:Se}
\end{center}
\end{figure}

\newpage
\clearpage

\newpage
\clearpage

\newpage
\clearpage

\newpage
\clearpage

\newpage
\clearpage

\newpage
\clearpage


\end{document}